\documentclass{evn2002}
\begin{document}
   \title{How is really decelerating the expansion of SN1993J?
          }

   \author{J. M. Marcaide\inst{1}
          \and
          A. Alberdi\inst{2}
          \and
          M. A. P\'erez-Torres\inst{3}
          \and
          J. C. Guirado\inst{1}
          \and
          L. Lara\inst{2,4}
          \and
          E. Ros\inst{5}
          \and
          P. J. Diamond\inst{6}
          \and
          F.~Mantovani\inst{3}
          \and
          I. I. Shapiro\inst{7}
          \and
          K. W. Weiler\inst{8}
          \and
          R. A. Preston\inst{9}
          \and
          R. T. Schilizzi\inst{10}
          \and
          R. A. Sramek\inst{11}
          \and
          C.~Trigilio\inst{12}
          \and
          S. D. Van Dyk\inst{13}
          \and
          A. R. Whitney\inst{14}
          }

   \institute{Departamento de Astronom\'{\i}a, Universitat de Val\`encia,
              E-46100 Burjassot, Spain
         \and
              Instituto de Astrof\'{\i}sica de Andaluc\'{\i}a, Apdo.\ Correos
              3004, E-18080 Granada, Spain
         \and
              Istituto di Radioastronomia, CNR, Via Gobetti 101,
              I-40129 Bologna, Italy
         \and
              Departamento de F\'{\i}sica Te\'orica y del Cosmos,
              Universidad de Granada, Avda.\ Severo Ochoa s/n,
              E-18071 Granada, Spain
         \and
              Max-Planck-Institut f\"ur Radioastronomie, Auf dem H\"ugel 69,
              D-53121 Bonn, Germany
         \and
              MERLIN/VLBI National Facility, Jodrell Bank Observatory,
              Macclesfield, Cheshire SK11 9DL, UK
         \and
              Harvard-Smithsonian Center for Astrophysics,
              60 Garden Street, Cambridge, MA 02138, US
         \and
              Remote Sensing Division, Naval Research Laboratory,
              Code 7213, Washington, DC 20375-5320, US
         \and
              Jet Propulsion Laboratory, California Institute of
              Technology, Pasadena, CA 91109, US
         \and
              Joint Institute for VLBI in Europe, Postbus 2,
              7990 AA Dwingeloo, The Netherlands
         \and
              National Radio Astronomy Observatory, Socorro, NM
              87801, US
         \and
              Istituto di Radioastronomia, CNR, I-96017 Noto , Italy
         \and
              Infrared Processing and Analysis Center, California
              Institute of Technology, Mail Code 100-22, Pasadena
              CA 91125, US
         \and
              Haystack Observatory, Massachusetts Institute of Technology,
              Westford, MA 01886, US
             }

   \abstract{
{SN~1993J is to date the radio supernova
whose evolution has been monitored in greatest detail and the one
which holds best promise for a comprehensive
theoretical-observational analysis. The shell-like radio structure
of SN~1993J has expanded in general accord with models of shock
excited emission, showing almost circular symmetry for over 8
years, except for a bright feature at the south-eastern region of
the shell that has been observed at every epoch. The spectrum of
SN1993J has flattened from $\alpha \simeq -1$ to $\alpha \simeq
-0.67$ $(S_{\nu }\propto \nu ^{\alpha })$. The decelerated
expansion can be modeled well with a single slope but apparently
better with two slopes. There are also intriguing hints of
structure in the expansion curve. The results by the two VLBI
groups carrying out this research show general agreement, but also
some differences. A comparison of the optical and VLBI results
about the details of the deceleration show some discrepancies.  }
   }

   \maketitle

\section{Introduction}
Radio emission from supernovae has been successfully modeled in
terms of the standard interaction model (SIM; Chevalier
\cite{che82}). This model considers fast-moving supernova ejecta
with steep density profiles ($\rho_\mathrm{ej} \sim
r^{-n}$) sweeping a circumstellar medium (CSM) of density profile $%
\rho_\mathrm{csm} \sim r^{-s}$, resulting in the formation of a high
energy-density shell. For $n > 5$, self-similar solutions exist,
and the shell radius evolves in time with a power law $R \sim
t^m$, where $t$ is the time since explosion and $m = (n-3)/(n-s)$
is the deceleration parameter. The radio emission is attributed to
synchrotron emission from relativistic electrons in the shell,
partially suppressed by external free-free absorption from thermal
electrons in the CSM.

\begin{figure}[htbp]
\vspace{240pt}
\includegraphics{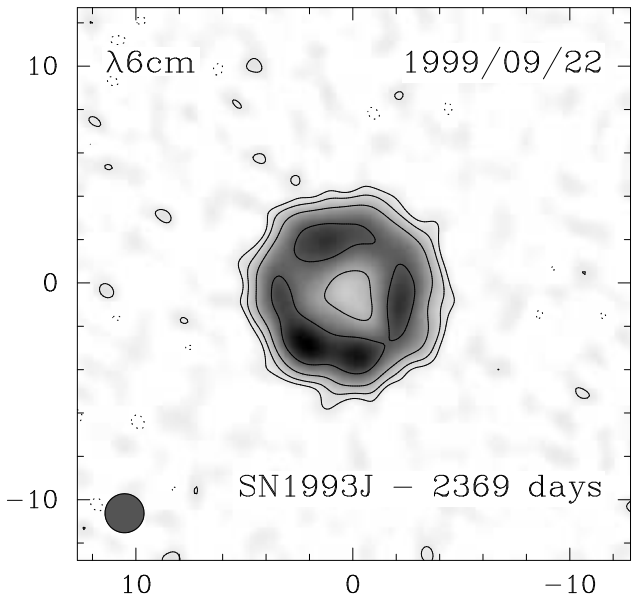} \caption{ 6-cm global VLBI image of
SN1993J in M81 corresponding to epoch 2369 days after the
explosion. The contours are spaced by  $2^{1/2}$ factors, from a
lower level of 160 \,$\mu$Jy\,beam$^{-1}$ to a brightness peak of
1939\,$\mu$Jy\,beam$^{-1}$. The convolving beam is circular, with
FWHM diameter of 1.8\,mas. The shell is still clearly defined, and
though its shape is not perfectly circular, it does not show any
evidence for strong asymmetries.} \label{fig:sep99-6cm-image}
\end{figure}
\begin{figure}[htbp]
\vspace{240pt} \includegraphics{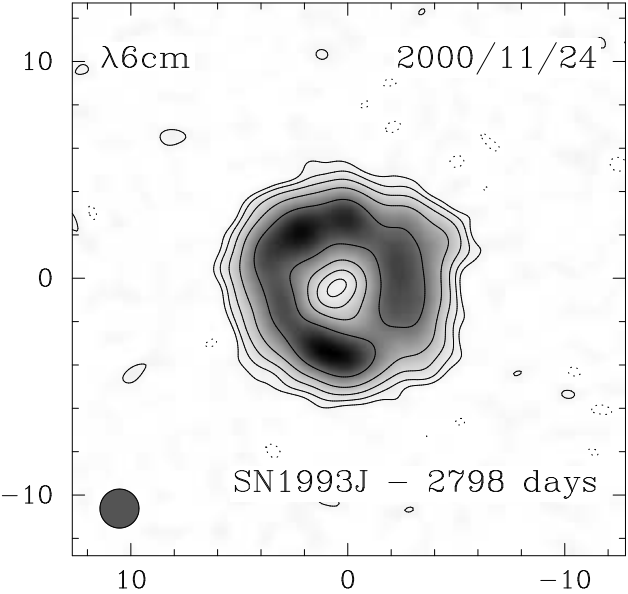} \caption{6-cm global VLBI
image of SN1993J in M81 corresponding to epoch 2798 days after the
explosion. The contours are spaced by  $2^{1/2}$ factors, from a
lower level of 47 \,$\mu$Jy\,beam$^{-1}$ to a brightness peak of
1327\,$\mu$Jy\,beam$^{-1}$. The convolving beam is circular, with
FWHM diameter of 1.8\,mas. The emission enhancements appear
somewhat different to Fig.~\ref{fig:sep99-6cm-image}}
\label{fig:nov00-6cm-image}
\end{figure}

Due to its proximity and its radio emission level, SN~1993J in M81
has offered an unprecedented occasion for VLBI studies. The
harvest of results includes: a) an initial source detection and
evolution by Marcaide et al. (\cite{mar94}) and Bartel et al.
(\cite{bar94}), respectively; b) the discovery of shell-like radio
structure (Marcaide et al. \cite{mar95a}); c) the first ``movie''
of an expanding supernova (Marcaide et al. \cite{mar95b}); d)
determinations of the deceleration in the expansion by Marcaide et
al. (\cite{mar97}) and Bartel et al. (\cite{bar00}); e) a
determination of the center of explosion of SN1993J relative to
the quasi-stationary core of M81 (Bietenholz et al. \cite{bie01}).
Results by the two groups carrying out this research show general
agreement, but also some differences like the determination of the
shell thickness ($\sim$30\% of the shell external radius by Marcaide et al.
(\cite{mar95b}); $\sim$20\% of the shell external radius by Bartel et al.
(\cite{bar00})). The angular expansion has so
far been rather smooth and circular and in accord with the SIM
model. However, the supernova shell has displayed, for every epoch
and wavelength, an enhancement of emission at its south-eastern
part (see, for instance, Fig.~\ref{fig:sep99-6cm-image} and
Fig. ~\ref{fig:nov00-6cm-image}, corresponding to supernova images
in September 1999 and November 2000, respectively, two of our last
observing epochs) probably related to the existence of small
anisotropies in the density distribution of the CSM. On the other
hand, our maps do not show yet any structures or protrusions
developing in the shell.

However, a closer look to the previous figures shows also
something not uncommon, namely, the changing enhancement of that
emission in the south-eastern part relative to other parts of the
structure. To be sure: the emission from the south-eastern part is
always enhanced, while the emission from other parts of the shell
seems to slightly come and go. Bartel et al. (\cite{bar00}) have
even suggested that there may be a cyclic pattern of changes in
shell azimuth. We do not have yet evidence of such thing from our
data. Even so, there may be at least two possibilities: (a) the
changing emission enhancements may be spurious due to, for
example, imperfect closure phases or artifacts associated with the
CLEAN algorithm, or (b) the changes are real and we should worry
to understand them. As said, we do not have evidence from our data
that the changes in different parts of the shell are any regular,
but perhaps our source sampling has not been appropriate. In any
case, the matter has to be systematically addressed with new
observations.

\begin{figure}[htbp]
\vspace{240pt} \includegraphics{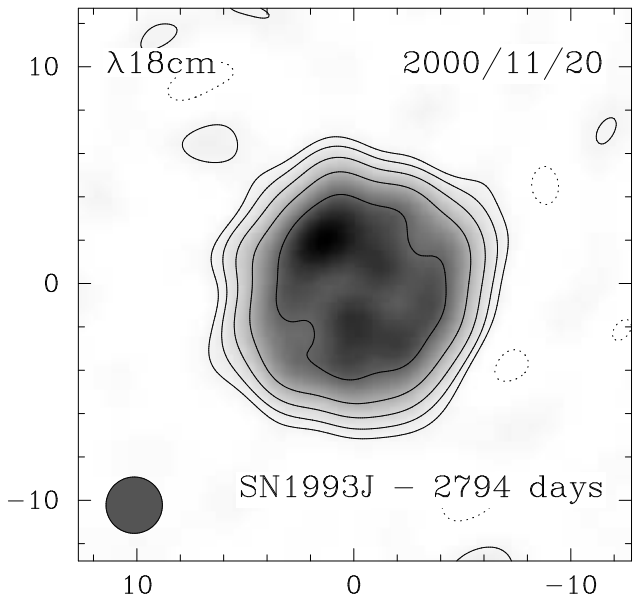} \caption{18-cm global
VLBI image of SN1993J in M81 corresponding to epoch 2369 days
after the explosion. The contours are spaced by  $2^{1/2}$
factors, from a lower level of 130 \,$\mu$Jy\,beam$^{-1}$ to a
brightness peak of 3430\,$\mu$Jy\,beam$^{-1}$. The convolving beam
is circular, with FWHM diameter of 2.6\,mas. The shell is not yet
clearly defined and its structure is reminiscent to that at 3.6-cm
prior to shell discovery from data 239 days after the explosion. }
\label{fig:18cm-image}
\end{figure}

To further complicate the picture, we show in
Fig.~\ref{fig:18cm-image} a  preliminary 18 cm image from November
2000. As expected, the shell is not yet clearly delineated at this
wavelength. However, is the emission enhancement location well
determined by the closure phases?

\section{Data Analysis}

\begin{figure}[htbp]
\vspace{180pt} \includegraphics{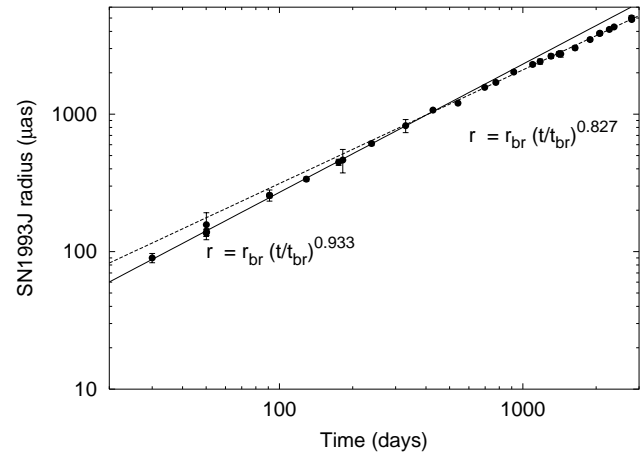}
\caption{ Weighted
least squares fit to the outer shell radius of SN1993J as a
function of the time since explosion, allowing for a
change in its deceleration rate. The VLBI data up to $t \leq t_{\mathrm{br}}$%
=403 days (where the solid and dashed lines in the figure cross
each other) can be well fitted by a power-law with index
$m_1$=0.933 (solid line), while
for $t \geq t_{\mathrm{br}}$ the best fit is given by power-law of index $%
m_2 $=0.827 (dashed line). The VLBI data for epochs below 180 days
are taken from Bartel et al. (\cite{bar00}). Note the logarithmic
scale.} \label{fig:deceleration-2log}
\end{figure}

From our VLBI measurements of SN~1993J at 3.6 and 6 cm, which
spanned from day 180 through day 1304 after explosion, we
determined a value for the deceleration parameter, $m=0.86\pm
0.02$ (Marcaide et al. \cite{mar97}). Bartel et al.
(\cite{bar00}), based on their VLBI observations at 1.3, 2, and
3.6 cm for the period from day 30 through day 1893, have claimed
that for 30$\leq $ t $\leq $ 306 days, $m_{1}=0.94\pm 0.02$, while
for 582$\leq t\leq $1893 days, $m_{2}=0.77\pm $ $0.01$. Adding the
results for 30$\leq $ t $\leq $ 175 days as given by Bartel et al.
(\cite{bar00}) to our data set, we have re-analyzed our data
through day 2798 allowing for a time-break in the expansion of
SN1993J (implying two deceleration parameters). We obtain the best
fit with the following parameters: $m_{1}=0.933\pm 0.019$,
$m_{2}=0.827\pm 0.008$, and $t_{\mathrm{br}}=403\pm 111$ days
(Fig.~\ref{fig:deceleration-2log}). This fit has a reduced
$\chi_{\nu } ^{2}$=0.51. Analyzing the same data set with one
deceleration parameter we obtain $m=0.87\pm 0.02$, with a reduced
$\chi _{\nu }^{2}$=1.37. Though the latter result is compatible
with our earlier one (Marcaide et al. \cite{mar97}), it has a
threefold larger reduced $\chi_{\nu }^{2}$. Then, although the
deceleration parameter for the early epochs is compatible with
that of Bartel et al. (\cite{bar00}) (not a surprise, since the
result is highly influenced by our use of the Bartel et al.
(\cite{bar00}) data for the period 30$\leq $ t $%
\leq $ 175 days when we have no data), the deceleration we obtain
for late epochs is significantly different. We should note here
that, for every epoch, the inferred source size depends on how the
map is constructed and how it is measured. Due to the finite size
of the VLBI beam, a positive bias is introduced in the size
estimate of each map. Marcaide et al. (\cite{mar97}) showed how to
account for this bias in order to obtain a correct estimate of the
deceleration parameter.

\begin{figure}[htbp]
\vspace{180pt} \includegraphics{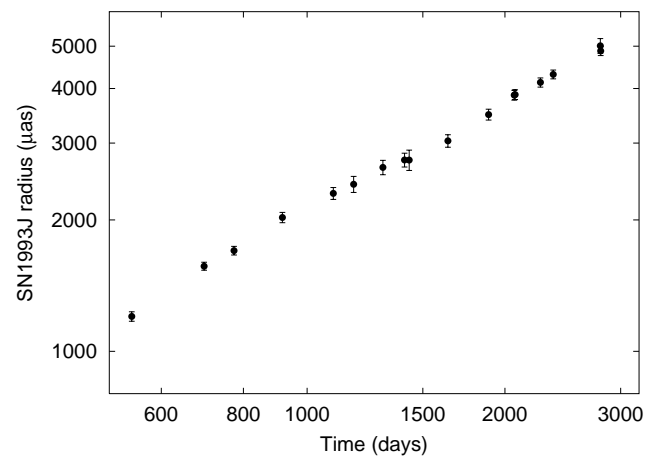} \caption{ The same data
as in Fig.~\ref{fig:deceleration-2log} but for the period from
$t$=500 days on, to enhance the fact that there might be a
modulation in the expansion of SN1993J between $t \sim$1200 and
$t\sim$2100 days. (See also the main text.) }
\label{fig:modulation-zoom}
\end{figure}

This picture gets even more complex if we consider the results of
the detailed analysis of early to late-time spectra of SN1993J
(Matheson et al. \cite{mat00}). Matheson et al. show that the
SN1993J velocity around 450-500 days, as given by the FWHM of the
H$_{\alpha }$ line, suffers a clear deceleration from about 23,000
km/s down to 17,000 km/s. However, from then on, and up to about
2400 days the velocity of the H$_{\alpha }$ line stays remarkably
constant at about 15,000 km/s. The VLBI data do not show such a
clear sign of deceleration (though our $t_{\mathrm{br}}$ of
$403\pm 111$ days is compatible with the epoch range given by
Matheson et al. \cite{mat00}). A fit to their data (equally
weighted) for epochs later than 500 days gives a deceleration
parameter, $m_\mathrm{Matheson}=0.87\pm 0.01$
[$v=v_{0}\,(t/t_{0})^{m-1}$]. This value, which is essentially our
$m=0.87\pm 0.02$ (single deceleration parameter), is closer to our
$m_{2}=0.83\pm 0.01$ than to $m_\mathrm{2,Bartel}=0.77\pm $ $0.01,$ as
reported by Bartel et al. (\cite{bar00}) for the period 582$\leq
t\leq $1893 days. The situation is not completely satisfactory.
Additionally, we are intrigued by an
apparent modulation in the expansion curve in the period 1200$\leq t\leq $%
2100 days (Fig.~\ref{fig:modulation-zoom}). This modulation could
be an artifact of the measurement process, but, so far, we have
been unable to trace it to anything or to relate it to our
measurement method. Mioduszewski et al. (\cite{mio01}) have
calculated the time evolution of the expansion parameter from
hydrodynamical simulations and have predicted the existence of a
time-dependent deceleration.

\begin{figure}[htbp]
\vspace{176pt} \includegraphics{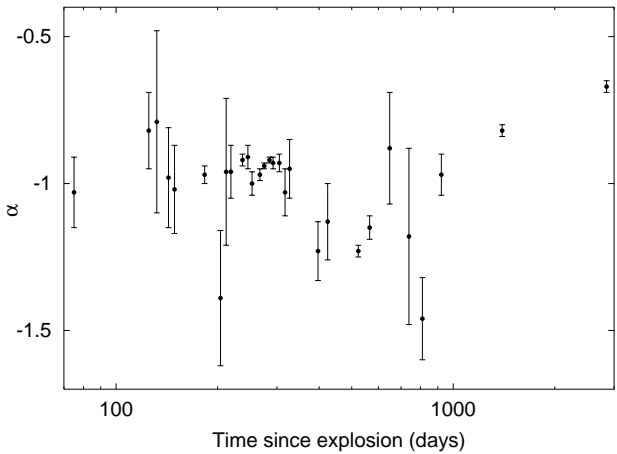}
\caption{{\protect\small Observed spectral index, $\alpha$, of
SN1993J from $t\sim$70 days up to $t$=2820 days, as obtained from
fitting at each epoch the available VLA data to a synchrotron
spectrum, partially suppressed by free-free absorption. After
$t\sim$1000 days the spectrum becomes clearly flatter. The data
used for the fits comes from measurementes made by
K.\ W.\ Weiler, 
except for the last data point (P\'erez-Torres et al.
\cite{per02}). }} \label{fig:alpha-change}
\end{figure}

The spectral index of SN1993J is also changing as shown in
Fig.~\ref{fig:alpha-change} (P\'{e}rez-Torres et al.
\cite{per02}). Quickly after explosion the supernova developed a
steep spectrum, characterized by an almost constant index $\alpha
$=-1.0 ($S_{\nu }\propto \nu ^{\alpha }$), but later, around day
1000, this spectrum became progressively less steep. Recent VLA
observations (P\'{e}rez-Torres et al. \cite{per02}), which include
observations at 90 cm, show that the spectral index around day
2820 has a value of $\alpha $=-0.67, that is, the spectrum of the
supernova is evolving towards a typical SN type II spectrum.
P\'{e}rez-Torres et al. (\cite{per02}) have found that a power-law
spectrum, free-free absorbed by an homogeneous --or clumpy, but not
a mixture of both-- distribution of ionized gas, yields the best
fit to the radio data.

\section{Future Prospects and Conclusions}

Over the coming years, we will try to carry on the VLBI monitoring
at 6 cm until the source becomes undetectable ($\sim 5 $ years),
and simultaneously observe at 18 cm. With these observations, we
will carefully monitor the deceleration, shell width, shell
brightness, and spectral index distribution. We will be able to
discern any possible dependence of those relevant parameters with
frequency and time. Such essential information will constitute the
input to our numerical simulation code (P\'{e}rez-Torres et al.
\cite{per01}) and hence it will strongly influence our ability to
characterize the physics involved. Also, our monitoring of the
structure of SN1993J should help to understand how the spectral
index changes in different parts of the structure.

Observations at 18cm are of outmost importance. Indeed, for
SN1993J it can be expected that years after an emission decline
(at 18 cm perhaps as long as 20 years) a nebular phase of
expansion of progressively increasing emission will appear as in
the case of SN1987A (Gaensler et al. \cite{gae97}) and SN1979C
(Montes et al.  \cite{mon00}). Then, the appropriate wavelengths
of observation will be the longest ones. Thus, it is relevant to
establish a long record of 18 cm observations of SN~1993J of the
maximum sensitivity now that the emission is optically thin and
the shell structure is becoming conspicuous at this wavelength.

\begin{acknowledgements}
This research was supported by the European Commission's IHP
Programme ``Access to Large-scale Facilities", under contract No.\
HPRI-CT-1999-00045 We acknowledge the support of the European
Union - Access to Research Infrastructure action of the Improving
Human Potential Programme.
M.A.P.T., P.J.D.\ and F.M.\ acknowledge partial support from the EC ICN
RadioNET (Contract No. HPRI-CT-1999-40003).
\end{acknowledgements}

\textbf{Final note:} {\footnotesize A comprehensive publication
(Marcaide et al. 2002) and a new movie on the expansion of SN1993J
are now being prepared. We will make soon the movie available to
everybody. A previous movie has been publicly available at the
JIVE and NRAO Image Galleries. It could also be retrieved by
anonymous ftp from jansky.uv.es/pub/jmm (sn93j-marcaide.avi, 33MB,
or sn93j-marcaide.avi.gz, 12MB)).}

\end{document}